\documentclass[sigconf]{acmart}

\copyrightyear{2024} 
\acmYear{2024} 
\setcopyright{rightsretained} 
\acmConference[ICSE '24]{2024 IEEE/ACM 46th International Conference on Software Engineering}{April 14--20, 2024}{Lisbon, Portugal}
\acmBooktitle{2024 IEEE/ACM 46th International Conference on Software Engineering (ICSE '24), April 14--20, 2024, Lisbon, Portugal}\acmDOI{10.1145/3597503.3623302}
\acmISBN{979-8-4007-0217-4/24/04}

\usepackage{xcolor}
\usepackage{multirow}
\usepackage{pifont}
\usepackage{sepnum}
\usepackage{xspace}
\usepackage[group-separator={,}]{siunitx}
\usepackage[most]{tcolorbox}
\usepackage{subcaption}
\usepackage{bigstrut}
\usepackage{makecell}
\usepackage{booktabs}
\usepackage{enumitem}

\newcommand{\cmark}{\ding{51}}%
\newcommand{\xmark}{\ding{55}}%
\newcommand*\rot{\rotatebox{90}}

\usepackage{colortbl}
\definecolor{gainsboro}{rgb}{0.86, 0.86, 0.86}
\newcommand{\clg}[1]{\cellcolor{gainsboro}{#1}}

\newcommand{\totalAttacks}{\num{127}\xspace}
\newcommand{\totalDamage}{\num{2331903028}\xspace}
\newcommand{\totalAttacksOutOfScope}{\num{95}\xspace}
\newcommand{\totalAttacksOutOfScopePercentage}{\num{75}\%\xspace}
\newcommand{\totalDamageOutOfScope}{\num{2060349987}\xspace}
\newcommand{\totalDamageOutOfScopePercentage}{\num{88}\%\xspace}
\newcommand{\totalAttacksInScope}{\num{32}\xspace}
\newcommand{\totalAttacksInScopePercentage}{\num{25}\%\xspace}
\newcommand{\totalDamageInScope}{\num{271553041}\xspace}
\newcommand{\totalDamageInScopePercentage}{\num{12}\%\xspace}
\newcommand{\totalLogic}{\num{47}\xspace}
\newcommand{\totalLogicPercentage}{\num{37}\%\xspace}
\newcommand{\totalLogicDamage}{\num{1116118649}\xspace}

\newcommand{\totalLogicInScope}{\num{75}\xspace}
\newcommand{\totalLogicInScopePercentage}{\num{59}\%\xspace}
\newcommand{\totalLogicInScopeDamage}{\num{1359921690}\xspace}
\newcommand{\totalLogicInScopeDamagePercentage}{\num{58}\%\xspace}

\newcommand{\point}[1]{\par\smallskip\noindent\textbf{#1.} }

\begin{document}
\sloppy

\title{Smart Contract and DeFi Security Tools: \\ Do They Meet the Needs of Practitioners?}

\author{Stefanos Chaliasos}
\affiliation{%
  \institution{Imperial College London}
  \country{United Kingdom}
}

\author{Marcos Antonios Charalambous}
\affiliation{%
  \institution{Imperial College London}
  \country{United Kingdom}
}

\author{Liyi Zhou}
\affiliation{%
  \institution{Imperial College London}
  \country{United Kingdom}
}

\author{Rafaila Galanopoulou}
\affiliation{%
  \institution{National and Kapodistrian University of Athens}
  \country{Greece}
}

\author{Arthur Gervais}
\affiliation{%
  \institution{University College London}
  \country{United Kingdom}
}

\author{Dimitris Mitropoulos}
\affiliation{%
  \institution{National and Kapodistrian University of Athens}
  \country{Greece}
}

\author{Benjamin Livshits}
\affiliation{%
  \institution{Imperial College London}
  \country{United Kingdom}
}

\renewcommand{\shortauthors}{Chaliasos et al.}

\begin{abstract}

The growth of the decentralized finance (DeFi) ecosystem built on blockchain technology and smart contracts has led to an increased demand for secure and reliable smart contract development. However, attacks targeting smart contracts are increasing, causing an estimated \$6.45 billion in financial losses. Researchers have proposed various automated security tools to detect vulnerabilities, but their real-world impact remains uncertain.

In this paper, we aim to shed light on the effectiveness of automated security tools in identifying vulnerabilities that can lead to high-profile attacks, and their overall usage within the industry. Our comprehensive study encompasses an evaluation of five SoTA automated security tools, an analysis of 127 high-impact real-world attacks resulting in \$2.3 billion in losses, and a survey of 49 developers and auditors working in leading DeFi protocols. Our findings reveal a stark reality: the tools could have prevented a mere 8\% of the attacks in our dataset, amounting to \$149 million out of the \$2.3 billion in losses. Notably, all preventable attacks were related to reentrancy vulnerabilities. Furthermore, practitioners distinguish logic-related bugs and protocol layer vulnerabilities as significant threats that are not adequately addressed by existing security tools. Our results emphasize the need to develop specialized tools catering to the distinct demands and expectations of developers and auditors. Further, our study highlights the necessity for continuous advancements in security tools to effectively tackle the ever-evolving challenges confronting the DeFi ecosystem.

\end{abstract}





\settopmatter{printfolios=true}
\maketitle

\section{Introduction}
The emergence of Ethereum and blockchains with smart contract
capabilities led to the development of decentralized applications (dapps),
opening up new possibilities for innovation.
The Decentralized Finance (DeFi) ecosystem,
which is built on these technologies,
has experienced significant growth since 2020,
with the total value locked (TVL) reaching an all-time
high of 180 billion USD on December 2021~\cite{defillama}.
Unfortunately,
this massive amount of value locked in DeFi
has also made them an attractive attack target.
Despite the efforts to write secure dapps,
attackers have successfully exploited vulnerable smart contracts
causing losses of 6.45 billion dollars~\cite{defillama},
underscoring the need for effective security measures.

Over the years,
researchers have dedicated tremendous efforts to
secure smart contracts by developing new techniques
and tools that identify 
vulnerabilities~\cite{kushwaha_2022,hu_zhang_liu_liu_yin_lu_lin_2021}.
Such techniques involve
static analysis~\cite{slither,securify,ethainter,vandal}, 
symbolic execution~\cite{mythril,oyente},
fuzzing~\cite{echidna,harvey,contractFuzzer,confuzzius},
formal verification~\cite{verx,verismart},
runtime verification~\cite{sereum},
and machine learning-based approaches~\cite{liu2018s,xfuzz}.
Despite these efforts,
high-profile attacks on smart contracts still persist.
To understand and evaluate these approaches,
researchers have conducted various studies.
Durieux et al.~\cite{smartbugs}
and Ren et al.~\cite{sc-evaluation-study-issta}
evaluated smart contract security tools,
while Perez and Livshits~\cite{expoited-sc}
assessed the high number of false positives of automated security tools.
Additionally,
Zhang et al.~\cite{demystifying-sc} performed a systematic
investigation to highlight missing vulnerability oracles.



Although there has been significant
research and focus on smart contract security,
it remains unclear how effective automated security tools
are against real-world exploits,
what impact these tools have on the industry,
and how they are utilized in developing and auditing smart contracts.
In this paper,
we aim to answer the following research questions.

\textbf{RQ1: Which vulnerability types can be detected by
automated security tools?}
How frequently do these vulnerabilities occur in real-world attacks?
What is the severity level of the vulnerabilities that
could have been detected by automated security tools in real-world attacks?
Finally, what types of vulnerabilities cannot be detected by
current automated security tools? (Section~\ref{sec:effectiveness})

\textbf{RQ2: To what extent can security tools be used to prevent
real-world high-profile attacks?}
What is the effectiveness of automated security tools against
each vulnerability category?
Which high-profile attacks could have been potentially avoided 
by using semi-automated security tools that require  user input?
(Section~\ref{sec:effectiveness})

\textbf{RQ3: What is the landscape of security tools used by
developers and auditors?}
To what extent do developers prefer open-source tools?
How prevalent are academic tools in practice?
What percentage of practitioners use semi-automated tools
that can prevent specific vulnerability types that are out-of-scope
for automated security tools?
How much time do auditors typically spend using
security tools during audits? (Section~\ref{sec:usage})
    
\textbf{RQ4: What are the key characteristics of security
tools that are prioritized by auditors and developers?}
Do practitioners weigh the trade-off between false positives
and false negatives when selecting security tools,
and how?
Additionally,
are ease of use,
documentation,
and report quality important factors when selecting
security tools for both developers and auditors?
(Section~\ref{sec:practitioners-perspective})
    
\textbf{RQ5: How effectively do security tools address
various classes of errors according to auditors and developers?}
Specifically, which types of errors are inadequately
covered by current security tools? Additionally,
what is the perception of auditors regarding the
usefulness of security tools?
(Section~\ref{sec:practitioners-perspective})

\point{Methodology} To address RQ1-RQ2,
we conducted an extensive empirical evaluation of
five automated security tools using a dataset of
127 high-impact real-world attacks.
In Section~\ref{sec:meth-empirical},
we describe the dataset,
the selection criteria we followed for the tools,
and our benchmarking process.
To answer RQ3-RQ5,
we conducted surveys with 49 developers and auditors
working in top DeFi protocols.
Our methodology for performing the surveys is presented in
Section~\ref{sec:meth-surveys}.

\point{Findings} Through our extensive analysis,
we have obtained the following findings
regarding the current state of security tools'
effectiveness and usage in the industry.

\textbf{RQ1:} Our empirical analysis revealed that the selected
automated security tools can identify 14 different types of vulnerabilities.
Among the attacks in our dataset, a total of~32 out of~127 exploits were associated with vulnerabilities in these~14 categories.
These~32 vulnerabilities resulted in a total damage
of approximately~271.5 million USD.
Notably, the top two types of vulnerabilities in the attack dataset involve concepts such as coding logic or sanity checks or on-chain oracle manipulation,
which in turn cannot be detected by current automated
security tools.

\textbf{RQ2:} The evaluation indicates that automated security tools
could have potentially prevented 11 out of 32 in-scope attacks,
resulting in a total loss of 149 million USD.
However,
security tools tend to generate numerous insignificant reports,
leading to a potentially overwhelming number of false positives.
All detected vulnerabilities were related to reentrancy,
highlighting the effectiveness of security tools against this type of vulnerability but also the inefficiency against other types.
Furthermore, 
our analysis indicates that existing security tools neglect protocol layer vulnerabilities.
Interestingly,
semi-automated tools could potentially prevent 47 attacks involving code logic absence, sanity checks, and logic errors.

\textbf{RQ3:} Our survey results show that developers
tend to use lightweight tools that can be easily integrated
into the development life cycle,
such as linters,
while auditors use more sophisticated tools with
greater bug-finding capabilities
(e.g.,
static analyzers).
The majority of developers (92\%) prefer open-source tools,
while over half of the participants reported using in-house security tools.
We found that academic tools used in research evaluations
and benchmark studies are not commonly used in practice.
Furthermore, about 59\% of developers and 48\% of auditors use tools
that can detect logic-related bugs,
which are often the root cause of high-impact attacks.
The majority of auditors (76\%) reported using security
tools for up to 20\% of their audit time.

\textbf{RQ4:} The results of our survey indicate that developers
prefer security tools with low false negative rates,
while auditors prefer tools with low false positive
rates since they are responsible for triaging reports.
In addition, auditors place a greater emphasis on the tool's
setup process and its bug-finding capabilities,
while developers prioritize tools that can be easily
integrated into their development workflows.
Both auditors and developers consider ease of use,
documentation,
and report quality to be important factors when
selecting security tools.

\textbf{RQ5:} Our findings reveal that both developers
and auditors consider logic-related bugs and oracle
manipulation vulnerabilities as significant threats
that are inadequately addressed by security tools.
They express the need for better support for these
types of vulnerabilities.
While the over half (52.4\%) of auditors find
security tools helpful for auditing,
a notable proportion (38.1\%) do not find them useful,
highlighting the need for further improvement in the
development and use of security tools for auditing purposes.


\textbf{Availability}: 
All the data and analysis from this study are accessible at 
\url{https://github.com/StefanosChaliasos/sc-defi-security/}.

\begin{figure*}[t]
    \centering
    \footnotesize
    \begin{tabular}{lcp{4cm}ccccccc}
    \toprule
    
    \bf Papers 
    & \bf Venue 
    & \bf Dataset 
    & \bf Vulnerabilities 
    & \bf Tools 
    & \multirow{2}{*}{\shortstack[l]{\bf Running\\\bf Tools}}
    & \multirow{2}{*}{\shortstack[l]{\bf Exploited\\\bf Contracts}}
    & \multirow{2}{*}{\shortstack[l]{\bf Measured\\\bf Impact}}
    & \multirow{2}{*}{\shortstack[l]{\bf Surveyed\\\bf Practitioners}} \\

    & & & & & & & & \\

    \midrule 
         Durieux et al.~\cite{smartbugs}
         & ICSE'20
         & 69 manually annotated vulnerable contracts and 49K on-chain contracts
         & 9 
         & 9 
         & \cmark
         & \xmark
         & \xmark
         & \xmark \\
         Ren et al.~\cite{sc-evaluation-study-issta}
         & ISSTA'21
         & 176 contracts retrieved from Github, EIPs, Academic papers
         & 6
         & 6 
         & \cmark
         & \xmark
         & \xmark
         & \xmark \\
         Perez et al.~\cite{expoited-sc}
         & USENIX SEC'21
         & 23K vulnerable contracts reported by six recent academic projects
         & 6
         & 6 
         & \cmark
         & 463
         & 1.7M USD
         & \xmark \\
         Zhang et al.~\cite{demystifying-sc}
         & ICSE'23
         & 516 from codearena and on-chain exploits
         & 17
         & 38
         & \xmark*
         & 54
         & \xmark
         & \xmark \\
         This work
         & 
         & 127 on-chain exploits
         & 39
         & 5
         & \cmark
         & 127
         & 2.3B USD
         & \cmark \\
    \bottomrule
    \end{tabular}%
    \caption{
Point-to-point comparison of related work on evaluating automated security tools. 
*: This paper categorizes bugs in machine unauditable bugs. Two tools, Slither and Oyente, were used not to measure false positives and false negatives, but to validate whether the tools were able to detect MUB bugs as defined in the study.
}
    \label{tab:related}%
\end{figure*}

\section{Background}
\label{sec:background}

The literature on the evaluation of automated security tools
for smart contracts has been primarily focused on assessing
their effectiveness by
constructing various benchmarks (see Figure~\ref{tab:related}).
Ferreira et al.~\cite{smartbugs-tool} developed Smartbugs,
an extendable evaluation framework that facilitates the integration and
comparison between multiple security tools that analyze EVM and Solidity smart contracts.
In~\cite{smartbugs},
the authors employed~$9$ automated analysis tools using two datasets;
one consisting of~$47$K contracts for consistency evaluation,
and the other one, $69$ annotated vulnerable contracts for precision evaluation.
Ren et al.~\cite{sc-evaluation-study-issta} proposed a comprehensive 4-step
evaluation process for minimizing bias in the assessment of automated tools.

Contrary to previous work,
our study aims to evaluate the real-world impact of automated security tools.
Perez and Livshits~\cite{expoited-sc} surveyed~23K smart contracts reported as vulnerable in 6 academic papers and found that only~$1.98$\% of them
had been exploited since deployment,
highlighting a potentially high number of false positives in existing techniques.
In contrast,
we focus on assessing automated tools' false negatives and gaining a deeper
understanding of their limitations.

Zhang et al.~\cite{demystifying-sc} performed a systematic investigation
of~$462$ defects reported in CodeArena audits and~$54$ exploits to study
the extent to which existing tools could detect them.
Our work takes a different approach by actually running the
tools against exploits and reporting both cases where the tools
have false negatives and cases where the tools lacked
appropriate oracles. 
Wan et al.~\cite{sc-practitioners-survey} surveyed
$156$ practitioners to understand their perceptions
and practices on smart contract security.
Our study on the other hand,
focuses on surveying dapp developers and auditors
to investigate how they use smart contract security tools.

In contrast to previous studies, this paper presents a mixed-methods investigation into the effectiveness and usage of security tools. The aim is to provide a comprehensive overview of the current status and offer valuable insights for researchers and practitioners to advance the state-of-the-art in smart contract and DeFi security.

\section{Methodology}
We provide an overview of the methods we employed to evaluate
the capability of current security tools to find real-world
vulnerabilities and understand practitioners' experience
when using such tools.
Specifically,
we describe the dataset containing real-world exploits,
the tool selection criteria,
and the benchmarking process.
Further,
we focus on the design of the surveys,
participant demographics,
and how we analyzed the results.

\subsection{Empirical Evaluation on Attacks}
\label{sec:meth-empirical}

\begin{figure}[tb]
    \centering
    \footnotesize
    \resizebox{1\linewidth}{!}{
    \begin{tabular}{lcccccccc}
    \toprule
    \multirow{2}{*}{\bf Vulnerability} 
    & \multirow{2}{*}{\bf Layer}
    & \multirow{2}{*}{\bf \#}
    & \multicolumn{5}{c}{\bf Tools} \\
    & & &
    \rot{Solhint~\cite{solhint}} & 
    \rot{Slither~\cite{slither}} & 
    \rot{Mythril~\cite{mythril}} & 
    \rot{ConFuzzius~\cite{confuzzius}} & 
    \rot{Oyente~\cite{oyente}} \\
    \midrule
Absence of coding logic or sanity check
& SC
& 42
& & & & & &\\
\clg{On-chain oracle manipulation}
& \clg{PRO}
& \clg{29}
& \clg{} & \clg{} & \clg{} & \clg{} & \clg{} \\
Reentrancy
& SC
& 13
& \ding{108} & \ding{108} & \ding{108} & \ding{108} & \ding{108} \\
\clg{Liquidity borrow, purchase, mint, deposit}
& \clg{PRO}
& \clg{10}
& \clg{} & \clg{} & \clg{} & \clg{} & \clg{} \\
Camouflage a token contract
& PRO
& 9
& & & & & \\
 \clg{Token standard incompatibility}
& \clg{PRO}
&  \clg{8}
& \clg{} & \clg{} & \clg{} & \clg{} & \clg{} \\
Function/State Visibility Error
& SC
& 8
& \ding{108} & & & & \\
\clg{Other unsafe DeFi protocol dependency}
& \clg{PRO}
& \clg{7}
& \clg{} & \clg{} & \clg{} & \clg{} & \clg{} \\
\makecell[l]{Other Inconsistent, improper or \\unprotected access control}
& SC
& 5
& \ding{108} & \ding{108} & \ding{108} & \ding{108} & \ding{108} \\
\clg{Logic Errors}
& \clg{SC}
& \clg{5}
& \clg{} & \clg{} & \clg{} & \clg{} & \clg{} \\
Unfair slippage protection
& PRO
& 4
& & & & & \\
\clg{Unfair liquidity providing}
& \clg{PRO}
& \clg{4}
& \clg{} & \clg{} & \clg{} & \clg{} & \clg{} \\
Direct call to untrusted contract
& SC
& 4
& & & \ding{108} & & \\
\clg{Other protocol vulnerabilities}
& \clg{PRO}
& \clg{3}
& \clg{} & \clg{} & \clg{} & \clg{} & \clg{} \\
Governance attack
& PRO
& 3
& & & & & \\
\clg{Transaction Order Dependence}
& \clg{PRO}
& \clg{2}
& \clg{} & \clg{} & \clg{\ding{108}} & \clg{\ding{108}} & \clg{\ding{108}} \\
Other coding mistakes
& SC
& 2
& & & & & \\
\clg{Delegatecall to Untrusted Callee}
& \clg{SC}
& \clg{2}
& \clg{} & \clg{\ding{108}} & \clg{\ding{108}} & \clg{\ding{108}} & \clg{} \\
Unsafe call to phantom function
& PRO
& 1
& & & & & \\
\clg{Improper asset locks or frozen asset}
& \clg{SC}
& \clg{1}
& \clg{} & \clg{\ding{108}} & \clg{} & \clg{\ding{108}} & \clg{} & \\
\makecell[l]{Other unfair or unsafe DeFi protocol \\ interaction}
& PRO
& 1
& & & & & \\
\clg{Camouflage a non-token contract}
& \clg{PRO}
& \clg{1}
& \clg{} & \clg{} & \clg{} & \clg{} & \clg{} \\
Weak Randomness
& PRO
& 0
& \ding{108} & \ding{108} & & & \\
\clg{Unhandled or mishandled exception}
& \clg{SC}
& \clg{0}
& \clg{\ding{108}} & \clg{\ding{108}} & \clg{\ding{108}} & \clg{\ding{108}} & \clg{\ding{108}} \\
Unbounded or gas costly operation
& SC
& 0
& &  & \ding{108} & & \\
\clg{Timestamp Dependence}
& \clg{PRO}
& \clg{0}
& \clg{\ding{108}} & \clg{\ding{108}} & \clg{\ding{108}} & \clg{\ding{108}} & \clg{\ding{108}} \\
Shadowing State Variables
& SC
& 0
& & \ding{108} & & & \\
\clg{Outdated compiler or solidity version}
& \clg{SC}
& \clg{0}
& \clg{\ding{108}} & \clg{\ding{108}} & \clg{} & \clg{} & \clg{\ding{108}} \\
Integer Overflow and Underflow
& SC
& 0
& & \ding{108} & \ding{108} & \ding{108} & \ding{108} \\


    \bottomrule
    \end{tabular}}
    \caption{Summary of vulnerability categories and the number of corresponding exploits in the Zhou et al. dataset~\cite{defisok}. \ding{108} indicates tool support for a corresponding vulnerability type. An empty cell indicates that a tool does not support the respective vulnerability. \textbf{SC}: Smart Contract Layer, \textbf{PRO}: Protocol Layer. We exclude vulnerability types that
    (1) the tools cannot support and (2) do not exist in the dataset.
Note that one exploit can be caused due to multiple vulnerabilities.}
    \label{tab:vulnerabilities}%
\end{figure}

\begin{figure}[t]
    \centering
    \footnotesize
    \begin{tabular}{lr}
    \toprule
    \textbf{Attacks} & \totalAttacks \\
    \textbf{Damage} & \totalDamage~\$ \\
    \textbf{Attacks out of selected tools' scope} & \totalAttacksOutOfScope~(\totalAttacksOutOfScopePercentage) \\
    \textbf{Damage out of selected tools' scope} & \totalDamageOutOfScope~\$~(\totalDamageOutOfScopePercentage) \\
    \textbf{Attacks in selected tools' scope} & \totalAttacksInScope~(\totalAttacksInScopePercentage) \\
    \textbf{Damage in selected tools' scope} & \totalDamageInScope~\$~(\totalDamageInScopePercentage) \\
    \bottomrule
    \end{tabular}
    \caption{Overall descriptive statistics of the analysed attacks.}
    \label{tab:scope}%
\end{figure}

\point{Dataset}
We use the dataset of DeFi attacks presented by Zhou et al.~\cite{defisok} as a basis for our analysis.
The dataset includes a comprehensive analysis and classification of 181
real-world, high-impact DeFi attacks.
Attack details involve underlying vulnerabilities in smart contracts, corresponding exploits, and monetary losses.
The vulnerabilities are categorized into five layers including
{\it Network}, {\it Consensus}, {\it Smart Contract},
{\it DeFi Protocol},
and {\it Auxiliary Service}.
Our work focuses on the Smart Contract and the DeFi Protocol layers,
because these are typically the layers where developer errors occur
and security tools focus
their analyses.
Hence,
we filtered out all vulnerabilities related to other layers.
This resulted in a dataset of~\totalAttacks~attacks.
Figure~\ref{tab:vulnerabilities} presents the vulnerability types as
reported in~\cite{defisok} while Figure~\ref{tab:scope} depicts
the total impact of the corresponding attacks.
Additionally,
we downloaded the source code~\footnote{via etherscan when available}
and bytecode of the smart contracts 
that were attack targets.

We chose this dataset because it reflects the real-world attacks that have occurred in the smart contract and DeFi ecosystem. While other related works~\cite{smartbugs,sc-evaluation-study-issta} have employed datasets of known vulnerable contracts or contracts with induced vulnerabilities, we believe that our selection of real-world attacks provides a more representative sample of the types of vulnerabilities smart contract developers and auditors should be aware of because they have led to major losses in deployed protocols. Furthermore, the contracts in the dataset have greater complexity than minimal examples, making reasoning about them more challenging.

\point{Tools Selection}
To select the security tools for our study,
we first conducted an advanced keyword search on
Google Scholar~\footnote{``Smart contract'', ``Smart contract security'', ``Ethereum'', ``ETH'', ``Ethereum Virtual Machine'', ``EVM'', ``EVM bytecode'', ``Solidity'', ``Ethereum automated analysis tools'', ``Blockchain'', ``Blockchain security'', ``Ethereum security'', ``Ethereum vulnerabilities'', ``DeFi'', ``Decentralized Finance''}
and followed references to identify additional tools.
We also searched for security tools in GitHub repositories.
The above process resulted in 75 tools.

Next, we applied a number of criteria to narrow down our selection.
Specifically,
we focused on (1) the availability of source code (51 tools),
(2) maintenance (14 tools)~\footnote{We define maintained tools as those that have had commits in the last year.},
(3) ability to run automatically without input (7 tools),
(4) popularity among practitioners
(e.g., prioritize tools with more GitHub stars and survey results), and
(5) repeated use in academic papers (i.e., higher reference count and usage in evaluations/comparisons).
We also included at least one tool based on the
following techniques:
linting, static analysis, fuzzing, and symbolic execution.
Notably,
focusing on tools that are based on different
analyses methods is an important dimension
of our study.

Based on the above criteria,
we ended up with the following tools:
{\it ConFuzzius}~\cite{confuzzius},
{\it Mythril}~\cite{mythril},
{\it Oyente}~\cite{oyente},
{\it Slither}~\cite{slither},
and {\it Solhint}~\cite{solhint}.
Solhint, Slither, and Mythril are widely recognized as the most popular and up-to-date linter, static analyzer, and symbolic executor, respectively.
ConFuzzius is the most updated fuzzer that meets our selection criteria.
Despite Oyente not being actively maintained,~\footnote{There is minimal support from the SmartBugs team fixing various errors, so it can still be used.}
we chose to include it in our analysis due to its status as one of the earliest academic tools, and its continued use in evaluations of numerous academic works~\cite{smartbugs,sc-evaluation-study-issta,maian,confuzzius,contractFuzzer, clairvoyance, contractWard, defectChecker, ethertrust, reentrancy2, sasc, scompile, securify, sfuzz, solidityCheck, verismart, verx, zeus}.

Figure~\ref{tab:vulnerabilities} depicts
the vulnerabilities that each selected tool can identify.
Note that the tools cannot detect every programming error related to
a vulnerability type.
For example,
in the case of
``\textit{Other Inconsistent, improper or unprotected access control}'',
Slither can only detect some of the bugs that can lead to
this defect type.
In the supplementary material, we provide a comprehensive overview of the tool selection process and a detailed mapping between tool vulnerabilities and the vulnerability categories of Zhou et al.~\cite{defisok}.

\begin{figure}[t]
\centering
\includegraphics[scale=0.35]{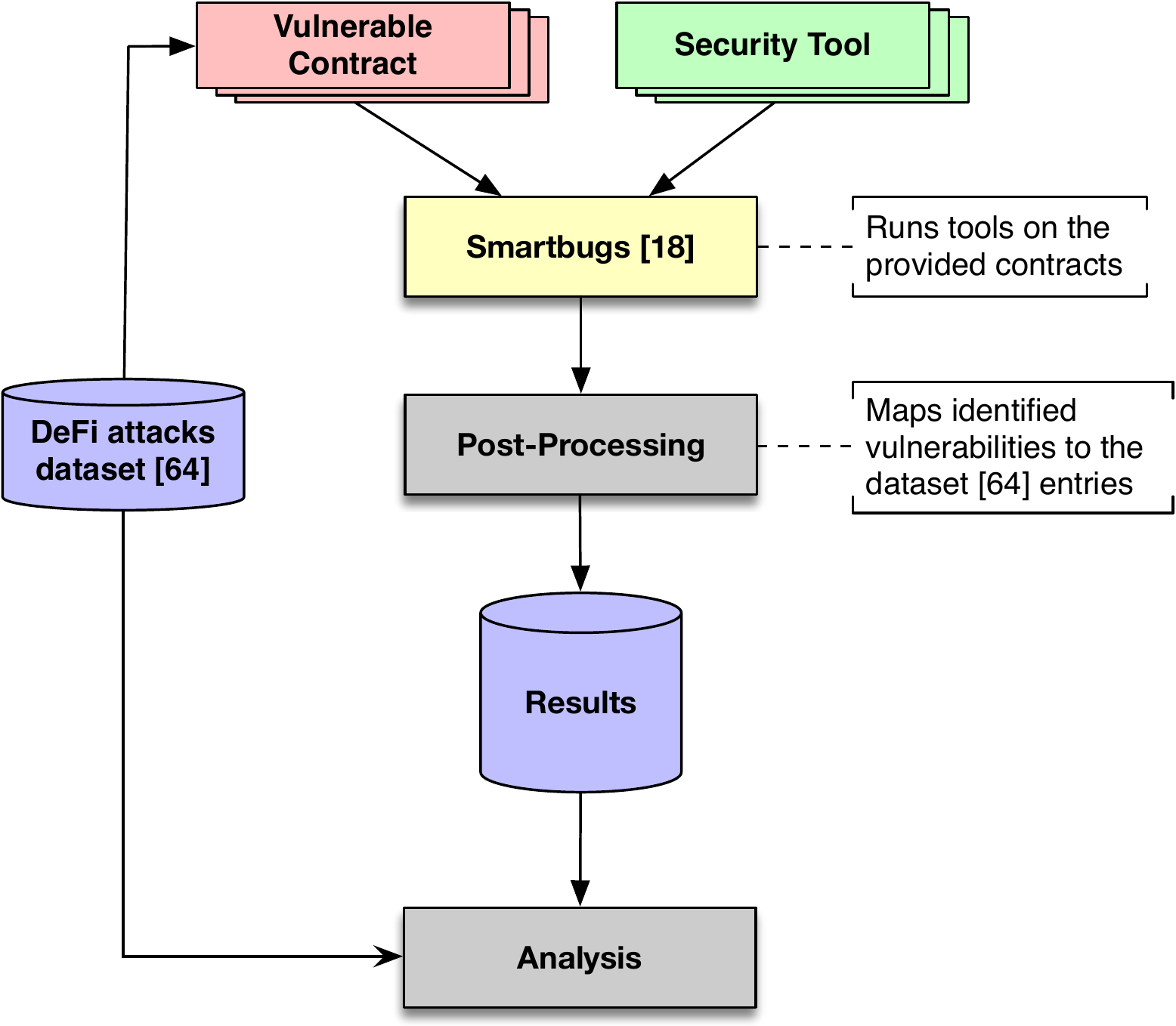}
\caption{Evaluating the effectiveness of security tools.}
\label{fig:arch}
\end{figure}

\point{Benchmarking}
Figure~\ref{fig:arch} summarizes our benchmarking approach.
To obtain results from the selected tools we utilized the SmartBugs framework~\cite{smartbugs-tool} (see also Section~\ref{sec:background}).~\footnote{We modified SmartBugs to ensure that the latest version of the selected tools was always used, all detectors capable of detecting security vulnerabilities were enabled, and a one-hour timeout per tool per contract was employed. When a tool could accept either source code or bytecode as input, we used both to evaluate its effectiveness.}
Next, we manually tracked all vulnerability types that each tool could detect and mapped them to the vulnerabilities of Figure~\ref{tab:vulnerabilities},
i.e., the vulnerabilities coming from the dataset.
We used a post-processing script to integrate this information
with the output of the SmartBugs framework and fed the data
into an SQLite database for further analysis.
Adding support for more tools is straightforward, as it only requires
including the tool in SmartBugs and provide a CSV
file that describes the mapping of the tool's detected
vulnerabilities to our toolchain.

After retrieving all results, we performed various sanity checks to verify
that the results were consistent.
In the case of a true positive,
two authors independently examined if the result is correct.
Solhint identified a number of defects of the following type: \textit{``Function/State Visibility Error''}, in five different exploited contracts.
However,
all cases were false alarms.
Finally,
we did not try to manually verify the rest of the results (i.e., potential false positives),
and we argue that most of the reports should have been either false 
positives or vulnerabilities that cannot be exploited in practice,
as the contracts in question had millions of USD in TVL,
and hence attackers would have had high motivation to attack them.

\subsection{Surveys}
\label{sec:meth-surveys}

\point{Protocol}
To better understand how developers and auditors
perceive and use security tools,
we conducted a survey campaign.
To do so,
we followed Kitchenham and Pfleeger’s guidelines~\cite{kitchenham}
(also used in similar studies~\cite{christakis16,sc-practitioners-survey}).
Further,
we employed best practices~\cite{SLMBZ13} to boost
practitioner participation.
Our survey was anonymous and we made all questions optional.
In addition,
we added an "other" option where possible to increase response rates.
Questions were divided into three categories:

\begin{enumerate}[leftmargin=*]
    \item \textbf{Demographics} to understand the respondents’ background.
    \item \textbf{Familiarity/usage of security tools during development and
    auditing} to assess if practitioners are well acquainted with security tools
    and how they use them.
    \item \textbf{Experience with security tools} to understand how satisfied practitioners are and how these tools can be improved.
\end{enumerate}

To fine-tune our campaign, we performed the following steps.
Two authors independently designed two slightly distinct surveys,
one for developers and one for auditors.
Then,
they converged on the questions that should be included in the
first versions of the surveys.
Moving forward,
the first round of the surveys took place with a set of
$N=3$ per survey where we asked the respondents to provide feedback.
After that first iteration,
we adjusted the questions and performed the same with $N=5$ per survey.
We used the feedback and responses to fine-tune the multiple-choice questions.

\point{Respondent selection and demographics}
Our aim was to focus our surveys on practitioners
with experience working on protocols with high TVL, which in turn,
are the main targets of attackers.
Instead of focusing on getting as many responses as possible,
we focused on obtaining high-quality responses.
Although this strategy might bias our results,
it was essential to focus on developers of top protocols
and auditors who assess such protocols to understand
the direct impact of security tools.

To recruit respondents,
we first contacted developers from the top 200 DeFi protocols,
as reported by Defillama~\cite{defillama}.~\footnote{Queried at 15/1/2023.}
For auditors,
we looked at the auditing companies with the most audit reports
for the top 200 protocols and contacted auditors from those companies.
We also contacted the top~100 auditors from Code4Arena~\cite{code4rena},
as independent auditors are also involved in auditing high-profile projects.
We received a total of~49 responses:
for the developers survey, out of the $266$ messages/emails sent, 
we successfully received $27$ responses, 
resulting in a response rate of $10$\%. 
Similarly, for the auditor survey, 
we received responses from $22$ out of the $132$ messages sent, 
corresponding to a response rate of $16$\%.
Figure~\ref{fig:dem},
presents an overview of the demographics of our survey participants.

\begin{figure}[t]
    \footnotesize
    \centering
    \resizebox{1\linewidth}{!}{
    \begin{tabular}{lrr}
    \toprule
     & \bf Developers & \bf Auditors  \\
    \midrule
    \underline{Years of experience in smart contract development/auditing} & &  \\
    \quad more than 5 & 8 (31\%) & 5 (23\%)  \\
    \quad 3-5 & 6 (23\%) & 3 (14\%) \\
    \quad 1-2 & 10 (39\%) & 6 (27\%)  \\
    \quad less than one & 2 (8\%) & 8 (36\%)  \\
    \underline{Organization's size} & &  \\
    \quad more than 250 & 0 (0\%) & 1 (5\%)  \\
    \quad 50-250 & 5 (19\%) & 5 (24\%) \\
    \quad 26-50 & 3 (12\%) & 1 (5\%)  \\
    \quad 6-25 & 15 (58\%) & 9 (43\%)  \\
    \quad 1-5 & 3 (12\%) & 0 (0\%)  \\
    \quad Independent & n/a  & 5 (24\%)  \\
    \underline{Main Targeted blockchains} & &  \\
    \quad Ethereum & 25 (93\%) & 22 (100\%) \\
    \quad Polygon & 15 (56\%) & 15 (68\%)  \\
    \quad Avalanche & 6 (22\%) & 10 (46\%) \\
    \quad Arbitrum & 11 (41\%) & 9 (41\%) \\
    \quad BSC & 7 (26\%) & 8 (36\%) \\
    \quad Fantom & 7 (26\%) & 11 (50\%) \\
    \quad Solana & 2 (7\%) & 3 (14\%) \\
    \quad Other & 6 (22\%) & 5 (23\%) \\
    \underline{Status of most mature dapp developed} & &  \\
    \quad Mainet & 25 (96.2\%) & n/a  \\
    \quad Development & 1 (3.8\%) & n/a  \\
    \midrule
    \quad Total & 27 & 22 \\
    \bottomrule
    \hline
\end{tabular}}
    \caption{Survey participant demographics.}
    \label{fig:dem}
\end{figure}

\point{Data analysis}
We analyzed the results based on question types.
For multiple-choice and Likert-scale questions,
we reported respondent percentages per option.
For open-ended questions,
we followed an inductive approach in which two authors
separately performed open card sorting and regularly
discussed emerging themes until an agreement was reached.
In the rest of this work,
we report percentages given the total responses to each question.

\section{Results}
\label{sec:exp}

In this section, we present the findings of our mixed-method investigation aimed at addressing our research questions.

\subsection{Effectiveness and Impact of Security Tools on Real-World Exploits}
\label{sec:effectiveness}

\begin{figure}[tb]
    \setlength{\tabcolsep}{3pt}
    \centering
    \footnotesize
    \begin{tabular}{lcccrr}
    \toprule
    \bf Tool (Version)
    & \bf Method
    & \multirow{2}{*}{\shortstack[l]{\bf Attacks\\\bf In Scope}}
    & \bf Detected
    & \multirow{2}{*}{\shortstack[l]{\bf Damage\\\bf In Scope}}
    & \multirow{2}{*}{\shortstack[l]{\bf Detected\\\bf Damage}} \\
    & & & & \\
    \midrule
    ConFuzzius (0.0.2) & Fuzzing & $22$ & $0$ & \$~\num{256393948} & \$~\num{0} \\
    Mythril (0.23.15) & SE & $24$ & $1$ & \$~\num{263104948} & \$~\num{25236849} \\
    Oyente (480e725)* & SE & $20$ & $0$ & \$~\num{247443948} & \$~\num{0} \\
    Slither (0.8.3) & SA & $20$ & $11$ & \$~\num{213793948} & \$~\num{149792690} \\
    Solhint (3.3.8) & Linting & $25$ & $0$ & \$~\num{213292041} & \$~\num{0} \\
    \hline
    \bf Total & &  $32$ & $11$ & \$~\num{271553041} & \$~\num{149792690} \bigstrut[t]\\
    \bottomrule
    \end{tabular}
    \vspace{-0.2cm}
    \caption{Tool effectiveness and damage that could have been prevented. SE: Symbolic Execution, SA: Static Analysis. * Used the forked version from: \url{https://github.com/smartbugs/oyente}.}
    \label{tab:effectiveness}%
    \vspace{-0.4cm}
\end{figure}

\begin{figure*}[t]
    \centering
    \footnotesize
    \begin{tabular}{|l|cc|cc|cc|cc|cc|ccc|}
    \hline
    \bf Vulnerability 
    & \multicolumn{2}{|c|}{\bf Slither}
    & \multicolumn{2}{|c|}{\bf Oyente}
    & \multicolumn{2}{|c|}{\bf ConFuzzius}
    & \multicolumn{2}{|c|}{\bf Mythril}
    & \multicolumn{2}{|c|}{\bf Solhint}
    & \multicolumn{3}{|c|}{\bf Total} \bigstrut[t]\\
    \hline\hline
    & D & ODI 
    & D & ODI 
    & D & ODI 
    & D & ODI
    & D & ODI 
    & TA & D & ODI \bigstrut[t] \\
    \hline
    Reentrancy & \num{11} & \num{69} & \num{0} & \num{0} & \num{0} & \num{1} & \num{1} & \num{18} & \num{0} & \num{11} & \num{13} & \num{11} & \num{71}\\
    Function/State Visibility Error & \xmark & \xmark & \xmark & \xmark & \xmark & \xmark & \xmark & \xmark & \num{0} & \num{62} & \num{8} & \num{0} & \num{62}\\
    Other inconsistent, improper or unprotected access control & \num{0} & \num{8} & \num{0} & \num{0} & \num{0} & \num{0} & \num{0} & \num{9} & \num{0} & \num{10} & \num{5} & \num{0} & \num{19}\\
    Direct call to untrusted contract & \xmark & \xmark & \xmark & \xmark & \xmark & \xmark & \num{0} & \num{0} & \xmark & \xmark & \num{4} & \num{0} & \num{0}\\
    Transaction Order Dependence & \xmark & \xmark & \num{0} & \num{1} & \num{0} & \num{0} & \num{0} & \num{0} & \xmark & \xmark & \num{2} & \num{0} & \num{1}\\
    Delegatecall to Untrusted Callee & \num{0} & \num{5} & \xmark & \xmark & \num{0} & \num{0} & \num{0} & \num{0} & \xmark & \xmark & \num{2} & \num{0} & \num{5}\\
    Improper asset locks or frozen asset & \num{0} & \num{8} & \xmark & \xmark & \num{0} & \num{0} & \xmark & \xmark & \xmark & \xmark & \num{1} & \num{0} & \num{20}\\
    Weak Randomness & \num{0} & \num{11} & \xmark & \xmark & \xmark & \xmark & \xmark & \xmark & \num{0} & \num{0} & \num{0} & \num{0} & \num{11}\\
    Unhandled or mishandled exception & \num{0} & \num{84} & \num{0} & \num{0} & \num{0} & \num{3} & \num{0} & \num{25} & \num{0} & \num{47} & \num{0} & \num{0} & \num{93}\\
    Unbounded or gas costly operation & \xmark & \xmark & \xmark & \xmark & \xmark & \xmark & \num{0} & \num{17} & \xmark & \xmark & \num{0} & \num{0} & \num{17}\\
    Timestamp Dependence & \num{0} & \num{46} & \num{0} & \num{0} & \num{0} & \num{2} & \num{0} & \num{46} & \num{0} & \num{55} & \num{0} & \num{0} & \num{69}\\
    Shadowing State Variables & \num{0} & \num{55} & \xmark & \xmark & \xmark & \xmark & \xmark & \xmark & \xmark & \xmark & \num{0} & \num{0} & \num{55}\\
    Outdated compiler or solidity version & \num{0} & \num{69} & \num{0} & \num{73} & \xmark & \xmark & \xmark & \xmark & \num{0} & \num{103} & \num{0} & \num{0} & \num{109}\\
    Integer Overflow and Underflow & \num{0} & \num{47} & \num{0} & \num{5} & \num{0} & \num{7} & \num{0} & \num{36} & \xmark & \xmark & \num{0} & \num{0} & \num{69}\\
    \hline
    \end{tabular}
    \vspace{-0.1cm}
    \caption{Summary of tool results. D (Detected). ODI (Other Detected Issues): other findings including false positives, defects that cannot be exploited (e.g. in protected functions),
    or exploitable defects not included in the dataset (i.e., not used in the attacks).
    TA (Total Attacks).}
    \label{tab:tools-results}%
    \vspace{-0.4cm}
\end{figure*}

Recently, automated security tools for detecting vulnerabilities in smart contracts have received increased attention. Previous studies~\cite{smartbugs,sc-evaluation-study-issta,expoited-sc} have evaluated their effectiveness by measuring recall and precision on datasets containing contracts sourced from blockchains (e.g., Ethereum) or manually crafted vulnerable contracts. Additionally, Zhang et al.~\cite{demystifying-sc} surveyed automated tools to determine their ability to detect various vulnerability categories. However, a key question that remains unanswered is the real-world impact of these tools, particularly in preventing significant exploits. To address this question, we conducted a comprehensive analysis of vulnerabilities in DeFi protocols that have led to significant exploits and assessed the effectiveness of automated security tools in preventing these exploits. Additionally, we quantified the potential funds that could have been saved by utilizing these tools.

\point{Automated tools scope}
Figure~\ref{tab:vulnerabilities} illustrates the scope of the selected security tools. We find that the automated security tools have oracles for the vulnerabilities that lead to the exploit for only~\totalAttacksInScopePercentage~of the~\totalAttacks~attacks studied. These attacks cause a total of 271 M USD in monetary losses, amounting to~\totalDamageInScopePercentage~of the total damage incurred by attacks in the dataset (c.f.\ Figure~\ref{tab:scope}). Notably, the automated security tools do not have oracles for detecting certain critical vulnerabilities, such as \textit{absence of code logic or sanity checks} and \textit{oracle manipulation}. Conversely, the tools tend to focus on vulnerabilities that do not appear to be frequently targeted by adversaries in high-profile attacks, such as \textit{integer overflows and underflows}, as well as \textit{unhandled or mishandled exceptions} (see Figure~\ref{tab:vulnerabilities}).

\point{Tool effectiveness on real-world vulnerabilities}
Out of~$32$ attacks that automated security tools can reason about the underlying vulnerabilities, only~$11$ of them could have been detected and potentially prevented if the tools were used (see Figure~\ref{tab:effectiveness}).~\footnote{Given that the tools were available in the time of the attack.}
Figure~\ref{tab:tools-results} depicts the results of the tools. Slither detects the most vulnerabilities, but it also reports many false positives~(FP). This can be detrimental to the usability of security tools as the number of reports that cannot lead to exploits may overwhelm users. Furthermore, our evaluation indicates that all tools detect vulnerabilities that were not utilized to exploit the assessed contracts, with static analysis and linting tools reporting a greater number of potential false alarms in comparison to other methods.

\point{Detecting different vulnerability types}
Notably, all of the~$11$ aforementioned attacks were caused by reentrancy vulnerabilities, suggesting that the focus on reentrancy by academic researchers~\cite{reentrancy1,reentrancy2,reentrancy3,reentrancy4} has led to the development of effective tools for this category.
Despite the effectiveness of these tools in detecting reentrancy vulnerabilities, there are still major issues. Of the five selected tools, only three were able to detect at least one vulnerability that led to a significant exploit. Additionally, 10 of the vulnerabilities could only be detected by Slither. 

Automated security tools (see Figure~\ref{tab:vulnerabilities}) are unable to detect ``\textit{Absence of coding logic or Sanity check}'' and ``\textit{Logic errors}.'' Thus, it is crucial to determine how many attacks could have been prevented by tools capable of detecting such errors, such as property-based fuzzers, formal verification, and model-checking tools. Notably, such tools could have potentially prevented \totalLogicPercentage~(\totalLogic/\totalAttacks) of the exploits in the dataset, amounting to \totalLogicDamage~USD in damage. When combining these tools with automated security tools, the total number of (potentially) preventable exploits in the dataset rises to \totalLogicInScope, accounting for \totalLogicInScopePercentage of the attacks and \totalLogicInScopeDamage~USD (\totalLogicInScopeDamagePercentage) of the total damage. 
Our results complement those of Zhang et al.~\cite{demystifying-sc}, who found that $79.5$\% of real-world bugs cannot be detected by automated tools alone. However, their research did not consider the effectiveness of semi-automated tools. Zhang et al.~\cite{demystifying-sc} also observed that logical errors often have generalized abstract models, indicating that human involvement could be crucial in constructing testing oracles. This finding is consistent with our preliminary findings. We leave it to future work to evaluate the practical effectiveness of semi-automated tools that can detect logic-related bugs and to assess the difficulty of writing specifications/properties for smart contracts that have been exploited.

\point{Potential preventable losses}
Our analysis shows that the total funds that could have been saved if selected tools were employed amount to~\num{149792690} USD, highlighting the importance of security tools in protecting smart contracts.

\begin{tcolorbox}
\textbf{Conclusions} for RQ~1, RQ~2
\begin{itemize}[leftmargin=*]
    \item 
In a subset of~$32$ attacks that automated security tools could have detected, only~$11$ of the exploited vulnerabilities were detected, highlighting a significant missed opportunity to enhance the security of smart contracts.
    \item 
All of the detected vulnerabilities were related to \textit{reentrancy}, indicating the effectiveness of the tools in detecting this type of vulnerability but also highlighting the inefficiency of automated tools in detecting other vulnerabilities. 
    \item
The top two types of vulnerabilities, absence of coding logic or sanity checks and on-chain oracle manipulation, cannot be detected by current automated security tools. Moreover, we observe that the majority of protocol-layer vulnerabilities are out of the scope of security tools.
\item
Semi-automated tools may be able to prevent 47 attacks that involve absence of code logic or sanity checks and logic errors.
    \item
The tools generate many insignificant reports, leading to a potentially overwhelming number of false positives.
    \item 
The total funds that could have been saved if the tools were employed are~\num{149792690} USD, underscoring the importance of security tools in preventing smart contract vulnerabilities.
\end{itemize}

\textbf{Call to action:} 
Security tools should focus on detecting vulnerabilities beyond reentrancy to be more effective in securing smart contracts and DeFi applications.
\end{tcolorbox}

\point{Discussion}
Despite almost a decade of research and development, automated security tools are still inefficient in detecting vulnerabilities in real-world contracts with high TVL, while reporting many potentially insignificant issues. Hence, further research is needed to improve the effectiveness and usability of these tools to better protect against financial losses due to vulnerabilities in smart contracts, while it is important to add support for more vulnerabilities.

\subsection{Familiarity and Usage of Security Tools}
\label{sec:usage}

In this section, we aim to explore the role of security tools in
the smart contract development lifecycle and DeFi audits, specifically focusing on how practitioners utilize these tools. To address this question, we survey both developers and auditors. In the following, we present the results of the surveys and analyze their implications for the development of secure dapps and effective DeFi audits.

\begin{figure}[t]
\centering
\includegraphics[width=0.45\textwidth]{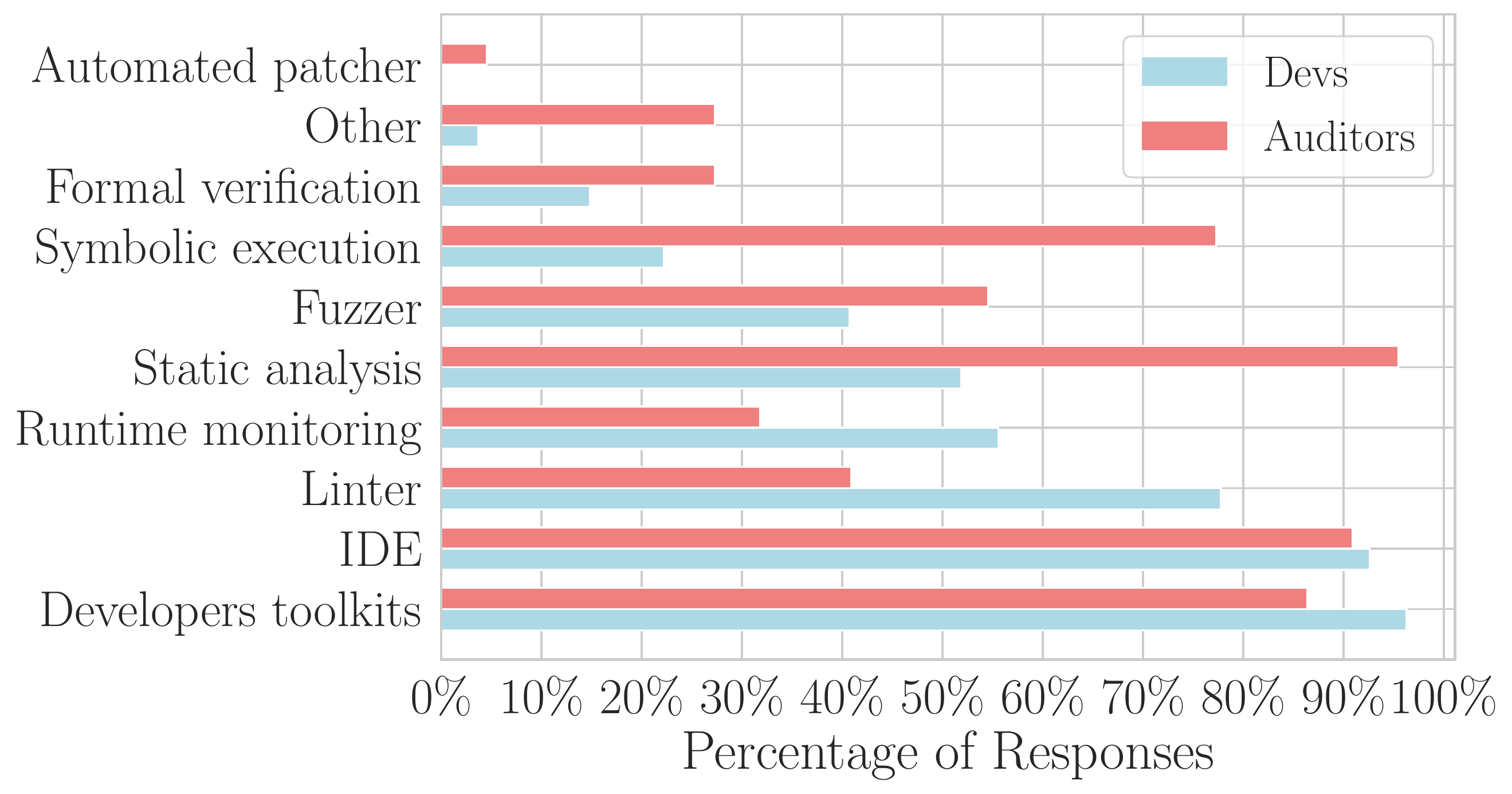}
\caption{Overall practitioner experience with different tool categories. \textit{Other} includes tools that employ more than one technique.}
\label{fig:experience}
\end{figure}

\point{Tool familiarity and usage for developers and auditors} 
Figure~\ref{fig:experience} illustrates the different types of tools that both developers and auditors have used. The category of tools that most practitioners have used is developer toolkits, followed by IDEs. These tools are primarily used for developing, deploying, debugging, and testing smart contracts. Interestingly, we observe that developers tend to favor lightweight tools such as linters, while auditors prefer tools with greater bug-finding capabilities, such as static analyzers and symbolic executors. Furthermore, developers have more experience using runtime monitoring tools, as most audits are performed on contracts before their deployment. Additionally, we found that developers have used an average of~4.5 different types of tools, while auditors have used an average of~5.3. 

\begin{figure}[tb]
\centering
\includegraphics[width=0.45\textwidth]{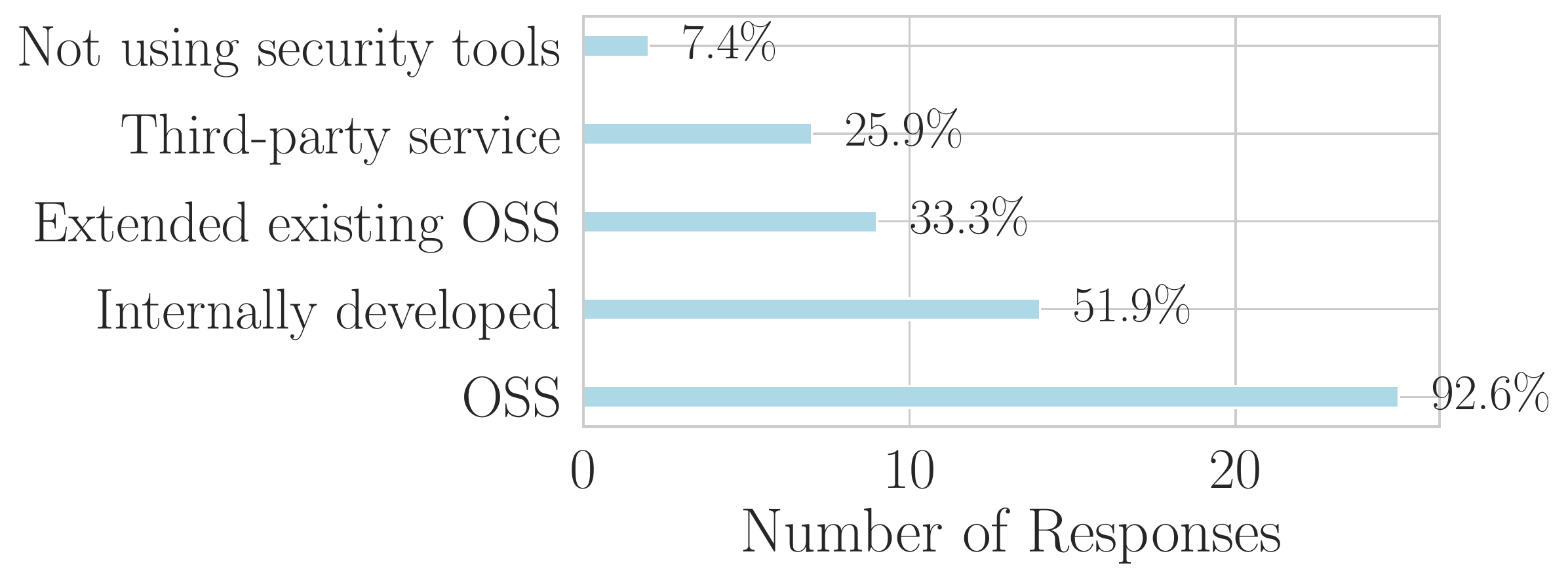}
\caption{Different tool types used during development.}
\label{fig:distirbution}
\end{figure}

\point{Reported tool usage}
We further investigated the types of security tools that practitioners use during development to secure decentralized applications (c.f.\ Figure~\ref{fig:distirbution}). Only two participants reported that they do not use any tool for this purpose. The majority~(92\%) of participants report that their organization utilizes open-source tools, and many invest effort into developing internal tools or extending existing open-source tools. Additionally,~25\% of participants reported that their organization uses third-party services typically provided by auditing firms. The prevalence of open-source tools highlights the importance of collaboration and community-driven efforts to improve security in the decentralized application ecosystem. Open-source tools have the potential to reach a wider audience and have a greater impact, ultimately leading to more secure and reliable decentralized applications.

\begin{figure}[t]
\centering
\includegraphics[width=0.45\textwidth]{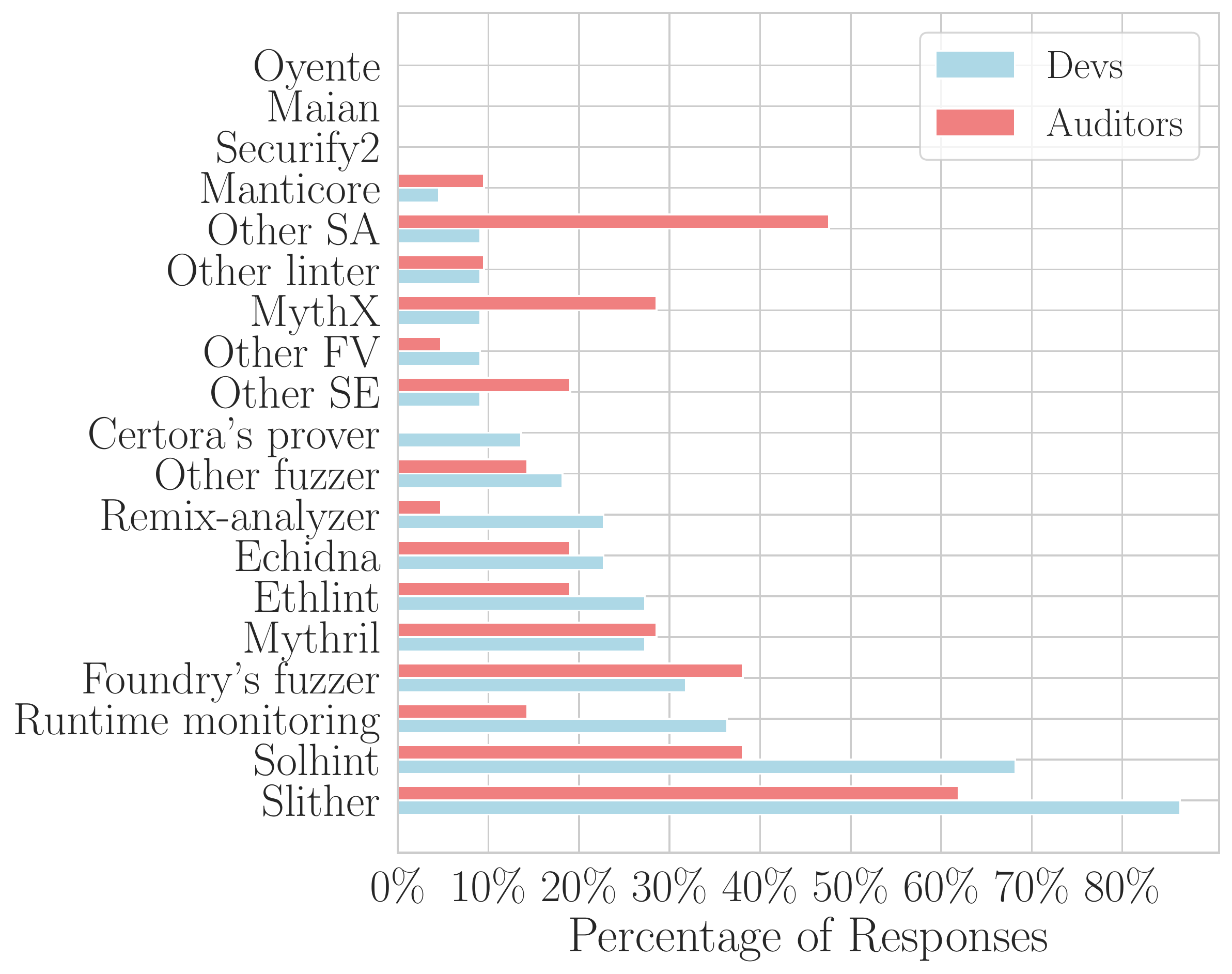}
    \caption{Security tools used by developers and auditors.}
\label{fig:tools}
\end{figure}

\point{Utility of specific tools during development and auditing}
Next, we explore which specific tools developers and auditors use during the development and auditing of dapps. Figure~\ref{fig:tools} displays the results of our investigation. The distribution of tool usage closely mirrors that of Figure~\ref{fig:experience}. It's worth noting that many auditors' responses included ``\textit{other}'' choices. This is because, in auditing companies, it is common for in-house security analysis tools to be developed and used in audits. Another noteworthy result is that various academic tools, such as Maian, Oyente, and Securify2, which are commonly used in scientific paper evaluations and benchmarking studies~\cite{sc-evaluation-study-issta,smartbugs,sc-study-icsme-20}, are not used in practice. This highlights the need for academia to adapt its comparisons and benchmarks to tools that are actually used by the community.

As we observed in Section~\ref{sec:effectiveness}, automated security tools cannot detect logic-related vulnerabilities. Thus, it is crucial to determine how many developers and auditors currently use tools capable of detecting such errors (i.e., formal verification and property-based fuzzing). Our surveys reveal that~59\% of developers and~48\% of auditors utilize at least one such tool.

\point{Property-based tests and application specifications}
As already mentioned, some tools require additional inputs, such as specifications of the smart contracts under test. Hence it is essential to understand who is responsible for providing these inputs. 40\% of the respondents indicated developers as responsible for writing specifications/property tests for semi-automated security tools, followed by 29\% for auditors and developers, 20\% for auditors, and 11\% of the respondents were unsure. As the effectiveness of such tools heavily relies on the quality of the provided inputs, we argue that both auditors and developers should participate in that process.

\point{Time spent on tool usage by auditors}
Another critical question to consider is how much time auditors spend running, fine-tuning, and validating the results of security tools. The results of our survey, indicate that the majority (76\%) of auditors spend a small proportion (between 0\%--20\%) of their time using such tools. 19\% spend between 21\% to 40\%, while 5\% spent 41\% to 60\%. This suggests that auditing is still primarily a manual effort. While there is certainly potential for tools to improve and automate certain aspects of the auditing process, it will continue to be predominantly a manual effort.

\begin{tcolorbox}
\textbf{Conclusions} for RQ~3.
\begin{itemize}[leftmargin=*]
    \item 
Overall we observe that developers tend to employ more lightweight tools, including linters, whereas auditors utilize tools with greater bug-finding capabilities (e.g., static analyzers). In addition, developers, use runtime monitoring tools more than auditors. 
    \item
Academic tools that appear in the context of research evaluations
and benchmark studies such as Oyente, are not used in practice.
    \item 
59\% of developers and~48\% of auditors utilize tools that can reveal logic-related bugs that are the root cause of many high-impact attacks.
    \item
The majority of auditors~(76\%) spent only up to~20\% of their time using security tools during audits, indicating that the auditing process is mainly a manual effort.
\end{itemize}

\textbf{Call to action.}
To bridge the gap between research and practice, researchers must consider three key factors. First, they should determine if the tools they create will be incorporated into development processes or employed during audits, focusing on prioritizing relevant features. Secondly, emphasizing the detection of vulnerability types that currently cannot be detected by existing security tools is vital. Finally, the evaluation of scientific papers should include benchmarks based on genuine real-world attack scenarios for more accurate and relevant results.
\end{tcolorbox}

\point{Discussion}
As different security tools may use varying techniques, with some being more resource-intensive than others, it is important to match tools appropriately to different stages of the development lifecycle. According to our survey results, developers tend to prefer tooling that can be used during the development process, such as linters, static analyzers, or after deployment, i.e. runtime monitoring tools. Therefore, it is crucial to develop tools that can be easily integrated into developers' daily routines. One such example is Foundry's property-based fuzzer~\cite{foundry}, which, despite being a relatively new tool, is already being utilized by a significant number of developers.

\subsection{What Makes Security Tools Valuable to Practitioners}
\label{sec:practitioners-perspective}

In this section, we aim to understand the factors that practitioners consider important when using security tools. Specifically, we explore the value that security tools provide in the context of detecting smart contract vulnerabilities and assess auditor satisfaction with the results generated by security tools. By examining these aspects, we can gain insight into what makes security tools valuable to practitioners and how they can be further improved to better serve the needs of the DeFi ecosystem.

\begin{figure}[t]
\centering
\begin{subfigure}[b]{0.45\textwidth}
   \includegraphics[width=1\linewidth]{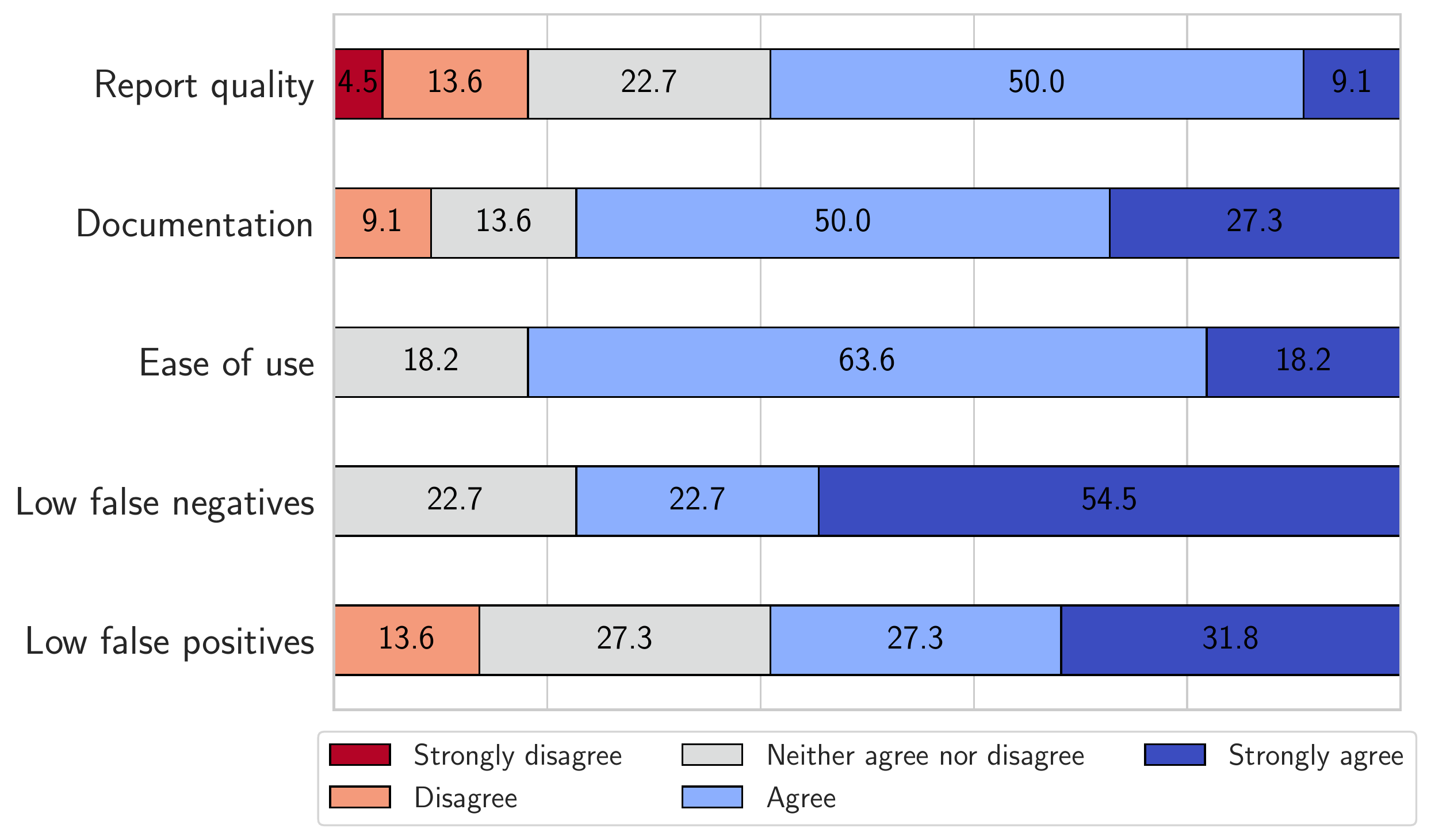}
   \caption{Auditors}
\end{subfigure}

\begin{subfigure}[b]{0.45\textwidth}
   \includegraphics[width=1\linewidth]{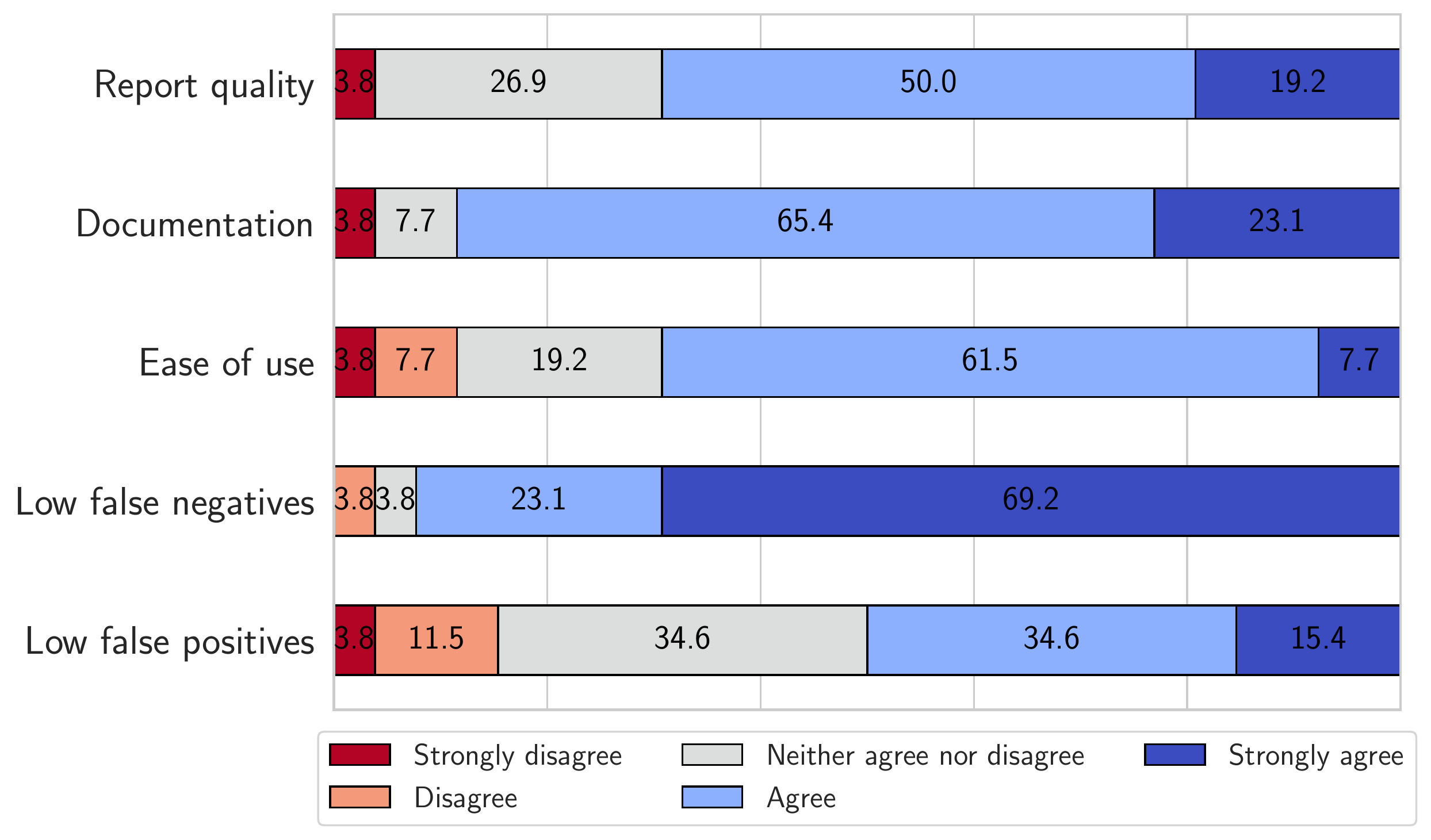}
   \caption{Developers}
\end{subfigure}
\caption{Importance of security tool characteristics.}
\label{fig:characteristics}
\end{figure}

\point{Importance of tools' characteristics}
Results from a Likert-based question on security tool characteristics are presented in Figure~\ref{fig:characteristics}. This survey question sought to understand the importance that both auditors and developers place on various aspects of security analysis. The results indicate that both groups consider all of the enumerated characteristics important, but there are some differences in the degree to which each characteristic is prioritized. For developers, low false negatives are perceived as more important mainly because they want reassurance that their applications are safe, whereas, for auditors, low false positives are considered to be more crucial, since it's their job to triage the reports. Additionally, ease of use is a bit more important for auditors, while some auditors do not place as much importance on report quality. Furthermore, we included one open-ended question about other factors that could positively or negatively affect the use of security tools. Many auditors emphasized the importance of tool setup, in addition, to ease of use. One participant highlighted the relation of time to configure / the severity of issues found. For developers, easy integration into the development life cycle (e.g., continuous integration), ease of customization, and the social aspect of other people using the tools and detecting important bugs in real-world applications were the most frequently mentioned factors. Overall, these findings highlight the diverse needs and priorities of practitioners when it comes to security tool features and underscore the importance of developing tools that meet a wide range of requirements.

\begin{figure}[t]
\centering
\begin{subfigure}[b]{0.45\textwidth}
   \includegraphics[width=1\linewidth]{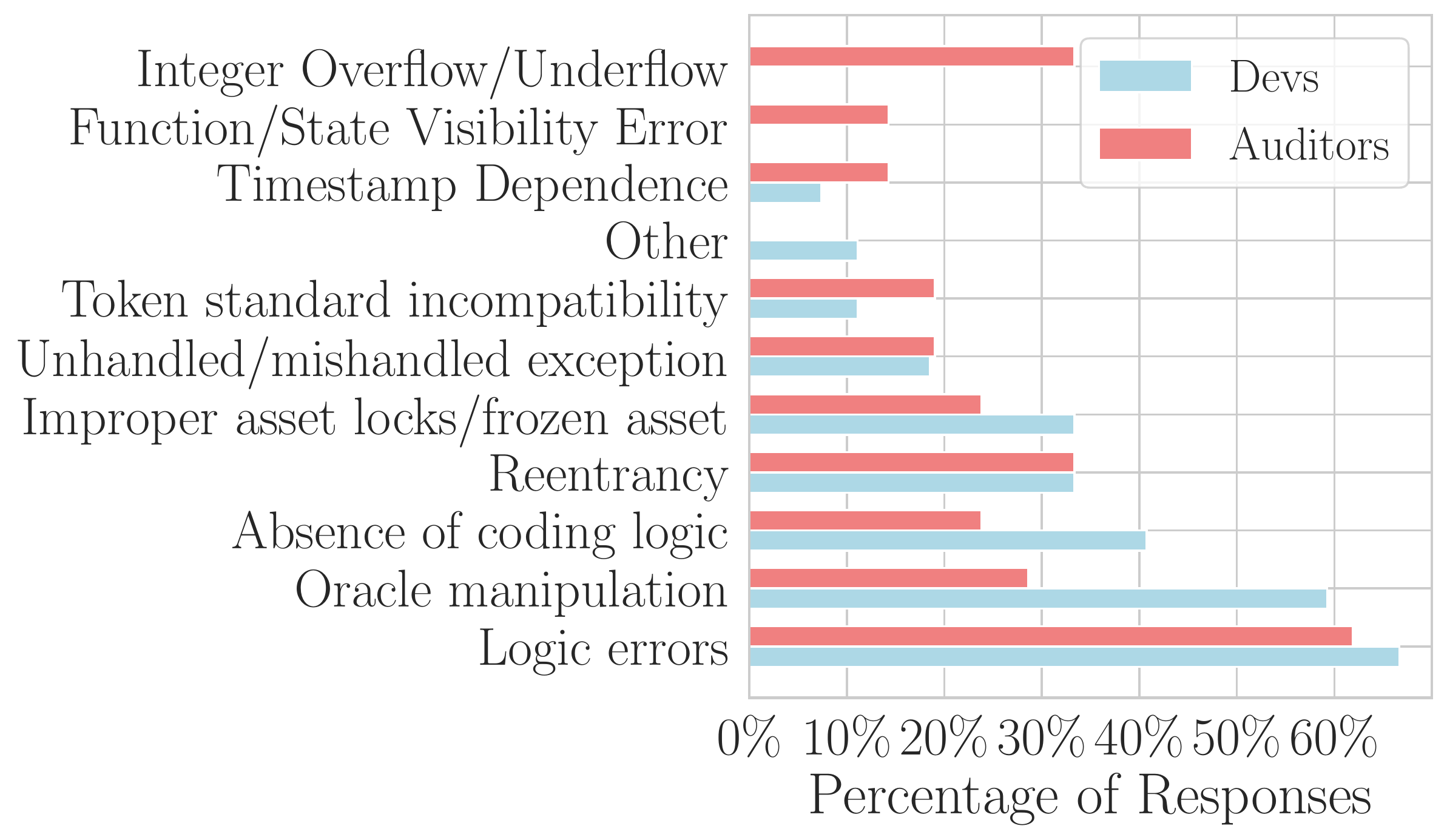}
   \caption{Opinions on the vulnerabilities that are
   difficult to identify (1) manually (auditors),
   and (2) during development (developers).}
   \label{fig:vulns}
\end{subfigure}

\begin{subfigure}[b]{0.45\textwidth}
   \includegraphics[width=1\linewidth]{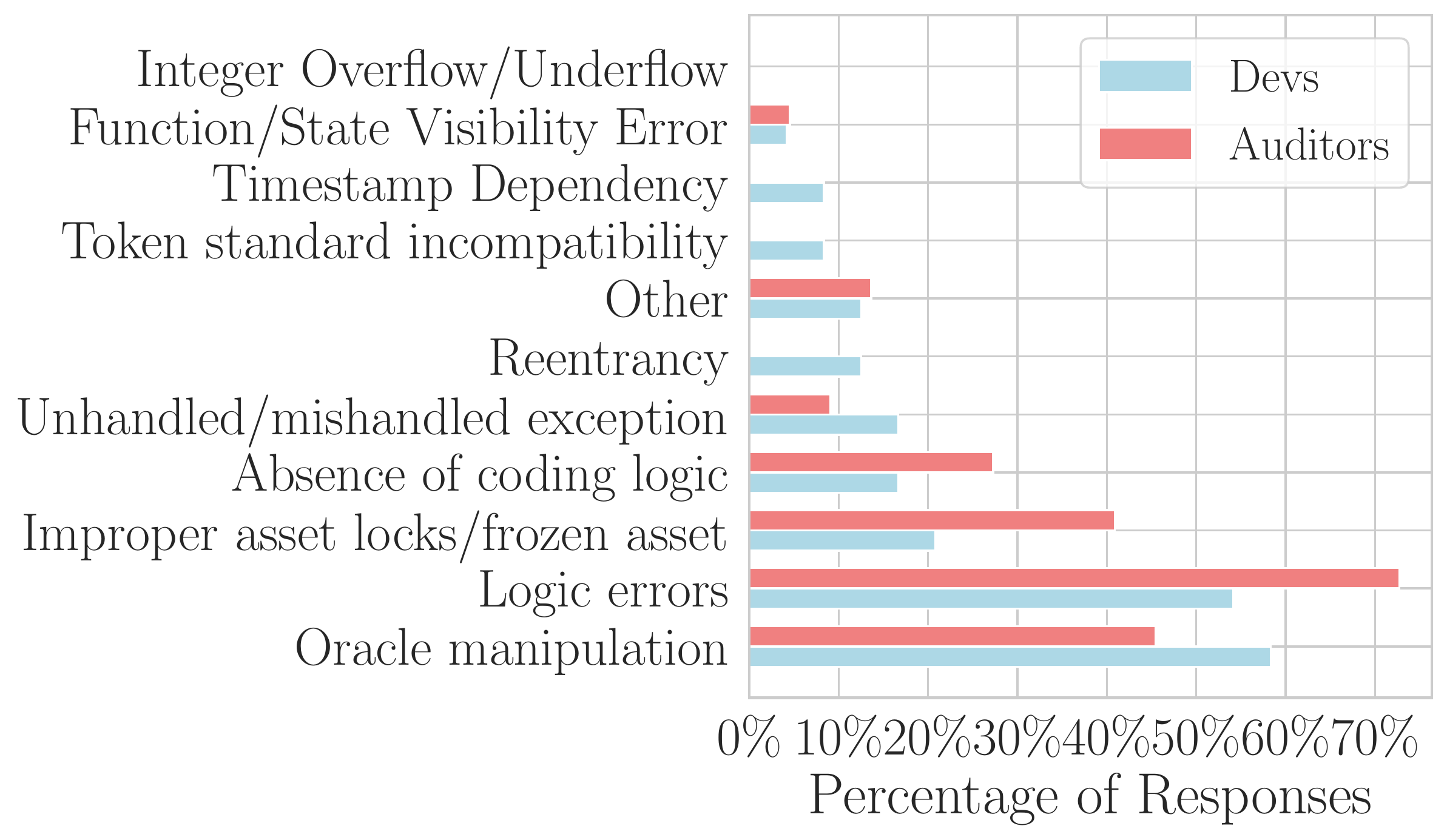}
   \caption{Opinions of developers and auditors on vulnerabilities that are both crucial and cannot be detected by security tools.}
   \label{fig:vulns-tools}
\end{subfigure}
\caption{Practitioner perspective on security tools and vulnerabilities.}
\end{figure}

\point{Exploring practitioners' perspectives on challenging vulnerabilities and available tooling for detecting such vulnerabilities} Figure~\ref{fig:vulns} sheds light on the most challenging vulnerabilities faced by both developers and auditors during the development and manual audit process, respectively. Developers identified logic errors, oracle manipulation, and absence of coding logic as the most difficult vulnerabilities to detect during development, which aligns with the state of most common defects in high-profile real-world attacks (see Figure~\ref{tab:vulnerabilities}). Auditors identified logic errors as the most challenging vulnerability to detect manually, followed by several vulnerabilities that existing tools have broad support for, such as integer overflows and reentrancy vulnerabilities, indicating that tools are indeed useful for identifying such bugs.

Regarding vulnerabilities that cannot be detected by automated security tools, both developers and auditors cited oracle manipulation and logic errors as the most challenging (c.f.\ Figure~\ref{fig:vulns-tools}). 
Additionally, both groups identified improper asset locks or frozen assets as a vulnerability that requires better support from tools. Overall, these findings emphasize the importance of developing more sophisticated security tools to detect crucial vulnerabilities that current tools may either not support or miss.

\begin{figure}[t]
    \centering
    \footnotesize
    \begin{tabular}{ccccc}
    \toprule
    1 & 2 & 3 & 4 & 5 \\
    \midrule
    9.5\% & 28.6\% & 33.3\% & 19\% & 9.5\% \\
    \bottomrule
    \end{tabular}
    \caption{Auditors' satisfaction ratings of security tools used for auditing, on a scale from 1 (not at all satisfied) to 5 (extremely satisfied).}
    \label{tab:auditors-sat}%
\end{figure}

\point{How auditors evaluate security tools for auditing smart contracts}
Our survey results indicate that a majority of participants found security tools helpful when auditing smart contracts, with 52.4\% rating them as 4 or 5 on a 5-point scale (c.f.\ Figure~\ref{tab:auditors-sat}). However, a significant portion of respondents (38.1\%) did not find security tools to be helpful or found them only somewhat helpful (rated 1-2). This suggests that there is still room for improvement in terms of the effectiveness and usability of security tools. In particular, participants highlighted the need for security tools to address more complex vulnerabilities that pose a greater threat to DeFi applications.

\begin{tcolorbox}
\textbf{Conclusions} for RQ~4 and RQ~5. 
\begin{itemize}[leftmargin=*]
    \item 
Developers prioritize low false negatives in security tools, while auditors prioritize low false positives in security tools (in comparison to developers), as it is their job to triage reports. Furthermore, auditors emphasize the importance of tool setup and bug-finding capabilities, while developers emphasize easy integration into the development lifecycle.
    \item
Both developers and auditors want better support for tooling related to logic-related and oracle manipulation vulnerabilities.
    \item
While~52.4\% of auditors find security tools helpful for auditing, a significant portion~(38.1\%) do not find them useful, highlighting the need for further improvement in the development and use of security tools in auditing.
\end{itemize}

\point{Call to action} 
Security tools should detect crucial vulnerabilities, such as logic-related and protocol-layer vulnerabilities (e.g., oracle-manipulation bugs), that can result in significant losses in practice. However, it is equally important for security tools to meet high usability and interoperability standards to be adopted by practitioners.
\end{tcolorbox}

\section{Discussion}

\subsection{Implications}

\point{Effectiveness, coverage, and need for manual inspection} Our analysis shows the limited effectiveness of automated security tools in detecting DeFi vulnerabilities. Figure 2 reveals that only 11 out of 32 (34\%) vulnerability types in our dataset were detected by the tools, emphasizing the insufficiency of current automated tools for comprehensive security assurance in DeFi ecosystems. Additionally, given the limited number of vulnerabilities covered by security tools (32/127), smart contract security relies heavily on manual inspections by designers, developers, or auditors. Our survey data indicates that only 59\% of developers and 48\% of auditors utilize tools capable of identifying logic-related errors, stressing the need for a holistic auditing approach combining automated tools and manual reviews.

\point{Emphasizing semi-automated tools for addressing critical vulnerabilities} Our findings point to the necessity of semi-automated security tools capable of detecting critical vulnerabilities in the smart contract ecosystem. Automated tools, while able to detect reentrancy vulnerabilities, fall short in covering logic-related bugs and protocol-layer vulnerabilities, such as oracle manipulation. Semi-automated tools, which incorporate user input to provide oracles for detecting security issues, present a promising solution. We encourage academia to focus on developing advanced tools that can effectively identify and prevent high-impact vulnerabilities, complementing practical tools already developed by the practitioner community~\cite{echidna,foundry}. For example, Liu and Li~\cite{invcon} have made progress in this direction by utilizing dynamic analysis techniques to identify invariants, which can subsequently be utilized as inputs for semi-automated tools.

\subsection{Threats to Validity}
We use a standard methodology~\cite{threats} to identify validity threats, which we mitigate where possible. 
This section discusses threats to internal, construct, and external validity for both the empirical analysis and the surveys.

\subsubsection{Empirical analysis}~\par

\point{Internal} One potential threat to internal validity is that the tools' results may be unsound. To mitigate this risk, we cross-checked the results and manually verified the essential findings. We also conducted sanity checks to ensure that the processing of the analysis results was correct. Another potential threat is that the dataset from Zhou et al.~\cite{defisok} may have incorrect data. To address this issue, we manually verified the important findings, while the dataset is open to the public for further verification. 

\point{Construct} A potential threat to construct validity is the setup of the tools used in the analysis. To mitigate this risk, we followed the documentation, the setting used in the tool papers, and ran the tools on both source code and bytecode when available. However, we note that ConFuzzius had the highest failure rate per contract, and while we attempted to mitigate any compilation errors, the tool failed in some cases because it could not deploy the targeted contract. ConFuzzius (and fuzzers in general) typically require more fine-tuning per execution, which is out of the scope of this work as we aimed to measure the out-of-shelf solutions available to practitioners with minimal setup. Another potential threat is mapping vulnerabilities from tools to the vulnerability types of Zhou et al.~\cite{defisok}. To address this issue, multiple authors performed the mapping independently, and we iterated over the mapping until reaching an agreement. We further consulted our mapping with the authors of~\cite{defisok}.

\point{External} A potential threat to external validity is the size of the DeFi attacks dataset. We used the most extensive dataset with attacks that have fine-grained information. Furthermore, we have automated the whole process, so adding more attacks to the analysis is straightforward.
Another potential threat is that we did not include all available tools in the analysis. In this work, we focused on tools most likely to be used by practitioners (c.f.\ Figure~\ref{fig:tools}). For each analysis technique, we selected the most well-established tool. We further included the most frequently used tool in academic paper evaluations (i.e., Oyente). Finally, we ran several not-maintained academic and industry tools (Securify2, SmartCheck, Conkas, Maian) and observed that their results did not change the paper's overall conclusions.

\subsubsection{Surveys}~\par

\point{Internal} Our survey responses may be subject
to a potential threat to internal validity,
as some respondents may not understand some of the questions well.
To reduce this risk,
we highlighted in the invitation message that all questions are optional,
and that they can skip any question that they do not understand.
Additionally, to mitigate this threat,
we designed our survey in an iterative fashion, as discussed
in Section~\ref{sec:meth-surveys}).

\point{External} Our goal is to survey developers
and practitioners that work on projects with high TVL
that are typically the targets of adversaries.
Hence, we meticulously selected who to invite and did not
share our surveys on social media or email lists to focus
on the quality of responses rather than quantity.

Furthermore, focusing on
developers and auditors who work on top protocols
and are more experienced with security tools
can pose an external validity threat,
because our results might not represent
the broader ecosystem.
Another risk involves the fact that a large percentage of the
participants work for the same organizations.
To mitigate this risk,
we sent up to three invites per organization.

\section{Related Work}
\point{Smart Contract Attacks and Security} 
To detect vulnerabilities in smart contracts, various tools using different techniques have been developed. Static analysis~\cite{securify, slither, ethainter, vandal, zeus} is one such approach, where the source code or bytecode is analyzed without execution. In contrast, dynamic analysis examines the smart contract while executing it. Fuzzing~\cite{echidna, contractFuzzer, harvey} is a testing technique where inputs are automatically generated to test the system's behavior. Symbolic execution~\cite{oyente, mythril, ethbmc, manticore} and formal verification~\cite{verismart,verx} are other well-known and frequently used techniques. However, formal verification typically requires users to provide specifications of intended behavior. In our study, we included one tool from each category that can be executed automatically, providing a comprehensive assessment of available solutions.

\point{DeFi Attacks and Security} DeFi attacks present unique challenges compared to those in traditional financial systems, primarily due to two key factors~\cite{qin2021cefi,werner2022sok}: \textit{(i)} the transparency in DeFi's application design, bytecode availability, and P2P transaction propagation; and \textit{(ii)} the composability of DeFi applications. 
Several studies have examined DeFi attacks, including Zhou et al.'s five-layered framework for incident categorization and evaluation~\cite{defisok}. Other significant works have focused on specific security issues. For instance, the Flash Boys paper~\cite{daian2020flash} was the first to explore the front-running issue, while Zhou et al. pioneered the study of sandwich attacks~\cite{zhou2021high}, which takes advantage of users' slippage settings in decentralised exchanges. 
DeFiRanger~\cite{wu2021defiranger} extracted DeFi actions and identified price oracle manipulation attacks using pattern matching. DeFiPoser~\cite{zhou2021just} employed SMT solvers to compose DeFi protocols, aiming to generate attacks. Collectively, these studies highlight the complexity and unique challenges posed by DeFi attacks. Additionally, our work underscores the limitations of traditional security tools that primarily focus on the smart contract layer neglecting the protocol layer.

\point{Surveys on smart contract vulnerabilities and security tools}
Atzei et al.~\cite{surveyattacks} performed the first survey of smart contract attacks. Chen et al.~\cite{surveyethereum} conducted a more comprehensive survey of 40 vulnerabilities, 29 attacks, and 51 defense locations and underlying causes, while Demolino et al.~\cite{delmolino16}, categorized bugs based on typical developer pitfalls. Harz et al. ~\cite{harz_knottenbelt_2018} investigated 10 smart contract verification tools, exhibiting various aspects of their security characteristics. Hu et al.~\cite{hu_zhang_liu_liu_yin_lu_lin_2021} assessed 39 analysis tools in terms of input type and methodology. Finally, Kushwaha et al.~\cite{kushwaha_2022} presented a comprehensive survey of 86 analysis tools, the most of any research publication and article, and examined their analysis approaches and tool type. In constrast to these studies, we focus on the real-world impact of security tools by evaluating them against high-profile attacks and surveying practitioners.

In a different spirit, Groce et al.~\cite{auditsfc} conducted an analysis of $23$ audits conducted by a prominent blockchain security company, employing a combination of automated tools and manual reviews. While security tools were utilized in $21$ of the audits, it is noteworthy that only $4$ out of $246$ identified vulnerabilities were explicitly detected by automated tools, specifically Slither. This finding supports the conclusion of our work that automated security tools require improvement to enhance their practical utility. Furthermore, despite the study being three years old, it identifies data validation (equivalent to \textit{absence of coding logic or sanity checks}) as the most common vulnerability within the audited contracts. This observation underscores the persistent threat posed by this category of bugs to the overall ecosystem.

\point{Surveys of program analysis and security tools}
Outside the realm of smart contracts security, Christakis et al.~\cite{christakis16} empirically investigate what appeals to practitioners the most about a program analyzer \cite{christakis16}, while~\cite{Smith20} evaluates the usability of security tools. Johnson et al.~\cite{johnson13} and Witschey et al.~\cite{witschey15}  explored why security tools are underused despite their benefits. On the contrary, in this work, we focus on how practitioners use security tools in the DeFi ecosystem. Finally, to the best of our knowledge, we are the first to survey auditors regarding security tool usage.

\section{Conclusions}
In conclusion, our evaluation of automated security tools, combined with surveys of developers and auditors, reveals that existing tools have limited effectiveness in detecting high-impact vulnerabilities, with only 8\% of the attacks in our dataset being detected by automated tools. This indicates that smart contract and DeFi security has not been fully addressed yet. While reentrancy vulnerabilities can be detected, the tools do not adequately address logic-related bugs and protocol-layer vulnerabilities.
We propose that researchers should prioritize the development of techniques that cover a wider range of vulnerabilities, including logic-related bugs, even if they partially require user input. Additionally, we suggest developing distinct tools for developers and auditors, as they have varying requirements regarding the capabilities of security tools.
We hope that our findings can provide valuable insights and guidance for practitioners and researchers working in this dynamic and challenging area.

\begin{acks}
We thank the anonymous reviewers for their constructive feedback. Further, we would like to express our sincere gratitude to all the participants who took the time to complete our surveys and provided invaluable feedback for our research. We also extend our thanks to Zhuo Zhang for the fruitful discussions on the impact of our results. This work has been partially supported by the European Union’s Horizon Europe research and innovation programme under grant agreement No 101070599.
\end{acks}

\bibliographystyle{ACM-Reference-Format}
\bibliography{main}

\end{document}